\documentclass[twocolumn,pra,aps,superscriptaddress]{revtex4-1}

\usepackage{graphicx}
\usepackage{amsfonts}
\usepackage{amsmath}
\usepackage{amssymb}
\usepackage{color}
\usepackage{hyperref}

\hypersetup{
colorlinks=false, linkcolor=black, urlcolor=black, pdfpagemode=none,
pdftitle={Cavity Cooling of an Ensemble Spin System},
pdfauthor={Christopher J. Wood, Troy W. Borneman},
  pdfsubject={Quantum information, quantum physics, spin physics}
  }

\definecolor{dred}{rgb}{.8,0.2,.2}
\definecolor{ddred}{rgb}{.8,0.5,.5}
\definecolor{dblue}{rgb}{.2,0.2,.8}
\newcommand{\bra}[1]{\mbox{$\langle #1|$}}
\newcommand{\ket}[1]{\mbox{$|#1\rangle$}}

\newcommand{\ketbra}[2]{\mbox{$|#1\rangle\langle #2|$}}

\newcommand{\tr}{\mbox{tr}}					% command for trace in math mode
\newcommand{\id}{\mathbb{I}}					% command for blackboard bold 1
\newcommand{\be}{\begin{equation}}				% Begin equation
\newcommand{\ee}{\end{equation}}				% End equation
\newcommand{\bea}{\begin{eqnarray}}			% Begin equation array
\newcommand{\eea}{\end{eqnarray}}				% End equation array

\def\mcal#1{{\mathcal #1}}
\def\1#1{{\bf #1}}
\def\2#1{{\cal #1}}
\def\3#1{{\sl #1}}
\def\4#1{{\tt #1}}
\def\5#1{{\sf #1}}
\def\6#1{{\mathfrak #1}}
\def\7#1{{\mathbb #1}}
\def\bb#1{{\mathbb #1}}

\begin{document}

% TITLE INFORMATION
\preprint{}
\title{Cavity Cooling of an Ensemble Spin System}
% CHRIS
\author{Christopher J. Wood}
\email{christopher.j.wood@uwaterloo.ca}
\affiliation{Institute for Quantum Computing, University of Waterloo, Waterloo, Ontario N2L 3G1, Canada}
\affiliation{Department of Physics and Astronomy, University of Waterloo, Waterloo, Ontario N2L 3G1, Canada}

% TROY
\author{Troy W. Borneman}
\email{tborneman@uwaterloo.ca}
\affiliation{Institute for Quantum Computing, University of Waterloo, Waterloo, Ontario N2L 3G1, Canada}
\affiliation{Department of Physics and Astronomy, University of Waterloo, Waterloo, Ontario N2L 3G1, Canada}

%CORY
\author{David G. Cory}
\affiliation{Institute for Quantum Computing, University of Waterloo, Waterloo, Ontario N2L 3G1, Canada}
\affiliation{Department of Chemistry, University of Waterloo, Waterloo, Ontario N2L 3G1, Canada}
\affiliation{Perimeter Institute for Theoretical Physics, Waterloo, Ontario N2L 2Y5, Canada}

\date{\today}

\begin{abstract}

We describe how sideband cooling techniques may be applied to large spin ensembles in magnetic resonance. Using the Tavis-Cummings model in the presence of a Rabi drive, we solve a Markovian master equation describing the joint spin-cavity dynamics to derive cooling rates as a function of ensemble size. Our calculations indicate that the coupled angular momentum subspaces of a spin ensemble containing roughly $10^{11}$ electron spins may be polarized in a time many orders of magnitude shorter than the typical thermal relaxation time. The described techniques should permit efficient removal of entropy for spin-based quantum information processors and fast polarization of spin samples. The proposed application of a standard technique in quantum optics to magnetic resonance also serves to reinforce the connection between the two fields, which has recently begun to be explored in further detail due to the development of hybrid designs for manufacturing noise-resilient quantum devices.

\end{abstract}

\maketitle

%=============================================================
%=============================================================
%		INTRODUCTION
%=============================================================
%=============================================================
\section{Introduction}
\label{sec:intro}

Efficient removal of entropy from a quantum system is essential for the development of robust quantum technologies and devices. High purity quantum states that may be quickly initialized and reset are necessary for the application of quantum error correcting codes to suppress and mitigate the effects of noise and errors that naturally occur in quantum information processors, sensors, and communication devices \cite{Terhal:2013a}. For spectroscopic applications, the signal-to-noise ratio increases significantly with state purity, allowing for the detection of small spin ensembles.

A spin ensemble may be naively prepared in a pure state by simply moving to low temperatures, where thermal fluctuations are not energetic enough to cause significant excitation out of the ground state. However, the required temperatures are often impractical to obtain or require sophisticated and expensive equipment. Additionally, the time required for the spin system to reach thermal equilibrium with the environment -- the energy relaxation time, $T_1$ -- often becomes very long at low temperatures, limiting the rate at which spin resets and signal averaging may be applied \cite{AbragamBook}. 

A variety of techniques for removing entropy from a quantum system are commonly used, including dynamic nuclear polarization (DNP) \cite{AbragamBook, Ramanathan:2008a}, algorithmic cooling \cite{Ryan:2008a}, optical pumping \cite{Weber:2010a}, laser cooling \cite{Wineland:1979a,Monroe:1995a,Vuletic:2000a}, and microwave cooling \cite{Valenzuela:2006a, Wallquist:2008a, Hauss:2008a}, among others. Recently, it was demonstrated that superconducting qubits may be prepared in an arbitrary pure state through sideband cooling by a high quality factor (high-Q) cavity \cite{Murch:2012a,Geerlings:2013a}. We discuss in this work how similar microwave cooling techniques should also be applicable to ensemble spin systems in magnetic resonance, despite the relatively small coupling between the cavity and a single spin. In particular, we present a theoretical model for how a high-Q resonator (cavity) may be used to actively drive each coupled angular momentum subspaces of a ensemble spin system to a state with purity equal to that of the cavity on a timescale significantly shorter than the thermal $T_1$ of the spins. Our model is motivated by recent studies that describe magnetic resonance in terms of quantum optics (for example, \cite{Jeener:2002a,Brahms:2010a,Hoult:2001a,Engelke:2010a,Tropp:2012a,Hahn:1997a}).

The ability to reduce the effective $T_1$ time of a spin ensemble by simply applying a detuned microwave drive provides an important tool for error correcting spin-based quantum information processors (for example \cite{Borneman:2012a,Cappellaro:2009a,Morton:2008a} and references therein), and should also find applications in spectroscopy by permitting faster signal averaging. These techniques may also find use in enhancing quantum memories for microwave photons based on coupling spin ensembles to superconducting devices (for example \cite{Staudt:2012a, Chiorescu:2010a, Schuster:2010a, Kubo:2010a, Xiang:2012a} and references therein).  

%=============================================================
%=============================================================
%		SYSTEM DESCRIPTION
%=============================================================
%=============================================================
\section{Mathematical Model}
\subsection{System Hamiltonian}
\label{sec:ham}
We consider an inductively driven ensemble of non-interacting spin-1/2 particles quantized in a large static magnetic field and magnetically coupled to a high-Q cavity. In the presence of the drive the spins interact with the cavity via coherent radiative processes and may be treated quantum mechanically as a single collective magnetic dipole coupled to the cavity \cite{Bonifacio:1970a}. In analogy to quantum optics, we describe the spin-cavity dynamics as being generated by the Tavis-Cummings (TC) Hamiltonian~\cite{Tavis1968,Tavis1969}. 
Assuming a linearly oscillating control field resonant with the Larmor frequency of the spins, the spin-cavity Hamiltonian is given by $H = H_0 + H_R(t) + H_I$, with 
\bea
H_0		&=& \omega_c a^\dagger a + \omega_s J_z	\\
H_R(t)	&=&  2\Omega_R \cos(\omega_s t)J_x		\\
H_I		&=& 2g(a^\dagger + a)J_x,  	\label{eqn:tc}
\eea
where $a^\dagger (a)$ are the creation (annihilation) operators describing the cavity, $\Omega_R$ is the strength of the drive field (Rabi frequency), $\omega_c$ is the resonant frequency of the cavity, $\omega_s$ is the Larmor resonance frequency the spins, and $g$ is the coupling strength of the cavity to a single spin in the ensemble in units of $\hbar=1$. Here we have used the notation that $J_\alpha\equiv \sum_{j=1}^{N_s} \sigma_\alpha^{(j)}/2$ are the total angular momentum spin operators for an ensemble of $N_s$ spins. 

The state-space $V$ of a spin-ensemble of $N_s$ identical spins may be written as the direct sum of coupled angular momentum subspaces $V=\bigoplus_{J={j_0}}^{N_s/2}V_J^{\oplus n_J}$ where $j_0 = 0 (1/2)$ if $N_s$ is even (odd). $V_J$ is the state space of a spin-$J$ particle with dimension $d_J = 2J+1$, and there are $n_J$ degenerate subspaces with the same total spin $J$~\cite{Sakurai}. Since the TC Hamiltonian has a global SU(2) symmetry it will not couple between subspaces in this representation. The largest subspace in this representation is called the Dicke subspace and consists of all totally symmetric states of the spin ensemble. It corresponds to a system with total angular momentum $J=N_s/2$. The TC Hamiltonian restricted to the Dicke subspace is known as the Dicke model~\cite{Dicke1954} and has been studied extensively for quantum optics (for a recent review see~\cite{Garraway2011}).

The eigenstates of $H_{0}$ are the tensor product of photon-number states for the cavity and spin states of collective angular momentum of each total-spin subspace in the $J_z$ direction: $\ket{n}_c\ket{J,m_z}_s$. Here $n=0,1,2,\hdots$, $m_z=-J,-J+1,\hdots,J-1,J$, and $J$ indexes the coupled angular momentum subspace $V_J$. The collective excitation number of the joint system for each subspace is given by $N_{ex}=a^\dagger a + (J_z+J)$. The interaction term $H_{I}$ commutes with $N_{ex}$, and hence preserves the total excitation number of the system. It drives transitions between the state $\ket{n}_c\ket{J,m_z}_s$ and states $\ket{n+1}_c\ket{J,m_z-1}_s$ and $\ket{n-1}_c\ket{J,m_z+1}_s$ at a rate of $\sqrt{n+1}\sqrt{J(J+1)-m_z(m_z-1)}$ and $\sqrt{n}\sqrt{J(J+1)-m_z(m_z+1)}$, respectively.

After moving into a rotating frame defined by $H_1=\omega_s (a^\dagger a + J_z)$, the spin-cavity Hamiltonian is transformed to
\bea
\widetilde{H}^{(1)}	
				&=& \delta\omega a^\dagger a + \Omega_R J_x + g(a^\dagger J_-+a J_+)
				\label{eqn:tcreduced}
\eea
where  $\delta\omega = \omega_c-\omega_s$ is the detuning of the drive from the cavity resonance frequency and we have made the standard rotating wave approximation (RWA) to remove any time-dependent terms in the Hamiltonian \cite{AbragamBook}. 

If we now move into the interaction frame of $H_2=\delta\omega a^\dagger a + \Omega_R J_x$, the Hamiltonian transforms to
\bea
\widetilde{H}^{(2)}(t) 
	&=& H_{0\Omega_R}(t) + H_{-\Omega_R}(t)+H_{+\Omega_R}(t)
	\label{eqn:hintfull}\\
H_{0\Omega_R}(t)	
	&=&  g\Big(
			e^{-i \delta\omega t}a +e^{i\delta\omega t}a^\dagger
		\Big)\,J_x 
					\nonumber	\\
H_{-\Omega_R}(t)	
	&=&\frac{i\,g}{2} \Big(
				e^{-i(\delta\omega-\Omega_R)t} a J_+^{(x)}
				- e^{i(\delta\omega-\Omega_R)t} a^\dagger J_-^{(x)} \Big)
				\nonumber\\
H_{+\Omega_R}(t)
	&=& \frac{i\,g}{2} \Big(
					e^{-i(\delta\omega+\Omega_R) t} a\, J_-^{(x)} 
					-e^{i(\delta\omega+\Omega_R)t} a^\dagger J_+^{(x)}
				\Big)
				\nonumber
\eea
where $J_\pm^{(x)}\equiv J_y\pm i J_z$ are the spin-ladder operators in the $x$-basis. 

In analogy to Hartmann-Hahn matching in magnetic resonance cross-relaxation experiments \cite{Bloembergen:1958a, Hartmann:1962a, Belthangady:2013} for $\delta\omega>0$ we may set the cavity detuning to be close to the Rabi frequency of the drive, so that $\Delta = \delta\omega-\Omega_R$ is small compared to $\delta\omega$. By making a second RWA in the interaction frame of $H_2$, the interaction Hamiltonian reduces to the $H_{-\Omega_R}$ flip-flop exchange interaction between the cavity and spins in the $x$-basis:
\bea
H_I(t)=\frac{i\,g}{2}\Big(e^{-i\Delta t}a\, J_+^{(x)}- e^{i\Delta t}\,a^\dagger J_-^{(x)}\Big).
	\label{eqn:hint}
\eea
This RWA is valid in the regime where the detuning and Rabi drive strength are large compared to the time scale, $t_c$, of interest $(\delta\omega,\Omega_R \gg 1/t_c$, (see~\ref{app:ham})\cite{supp_mat}). From here we will drop the $(x)$ superscript and just note that we are working in the $J_x$ eigenbasis.

Isolating the spin-cavity exchange interaction allows efficient energy transfer between the two systems, permitting them to relax to a joint equilibrium state in the interaction frame of the control field. The coherent enhancement of the ensemble spin-cavity coupling -- similar to the enhancement of the vacuum Rabi frequency for atomic ensembles, but not restricted to the single-excitation manifold \cite{Yamamotobook} -- enhances spin polarization at a rate that may exceed the thermal relaxation rate. 

We note that the spin-cavity exchange coupling also exists in the absence of the Rabi drive, and theoretically permits cooling of the spin system by matching the resonance frequency of the spin system to the cavity resonance. However, this process is thermally driven, and thus corresponds to a set of incoherent radiative processes that may not be described by a single Hamiltonian \cite{Bonifacio:1970a}. This Purcell effect in magnetic resonance systems has been previously noted and is normally small enough to be neglected~\cite{Purcell1946,Mollow:1969a}. 

%=============================================================
%=============================================================
%		MASTER EQUATION
%=============================================================
%=============================================================
\subsection{Master Equation for the spin-ensemble under cavity dissipation}
\label{sec:me}
 
To model the cavity-induced cooling of the spin system we use an open quantum system description of the cavity and spin ensemble. The joint spin-cavity dynamics may be modelled using the time-convolutionless (TCL) master equation formalism~\cite{Breuer2002}, allowing the derivation of an effective dissipator acting on the spin ensemble alone. Since the spin-subspaces $V_J$ are not coupled by the TC-Hamiltonian, the following derivation holds for all values of $J$ in the state-space factorization.

The evolution of the spin-cavity system is described by the Lindblad master equation
\be
\frac{d}{dt} \rho(t) = \2 L_I(t)\rho(t) + \2 D_c\rho(t)
\ee
where $\2 L_I$ is the super operator $\2 L_I(t)\rho = -i [H_I(t), \rho]$ describing evolution under the interaction Hamiltonian \eqref{eqn:hint}, and $\2 D_c$ is a dissipator describing the quality factor of the cavity phenomenologically as a photon amplitude damping channel\cite{Agarwal1974}:
\be
\2 D_c =\frac{\kappa}{2}\Big( (1+\overline n)\,\2 D[a] +\overline{n}\,\2 D[a^\dagger]\Big),
\label{eqn:diss}
\ee
where $\2 D[A](\rho) = 2\,A\, \rho\, A^\dagger-\{A^\dagger A, \rho\}$, $\overline{n} = \tr[a^\dagger a \rho_{eq}]$ characterizes the temperature of the bath, and $\kappa$ is the cavity dissipation rate ($\propto 1/Q$). The expectation value of the number operator at equilibrium is related to the temperature, $T_c$, of the bath by
\be
\overline{n} = \left(e^{\omega_c/k_B T}-1\right)^{-1} \Leftrightarrow T_c = \frac{\omega_c}{k_B}\left[\ln\left(\frac{1+\overline{n}}{\overline{n}}\right)\right]^{-1}
\label{eqn:ctemp}
\ee
where $k_B$ is the Boltzmann constant. 

The reduced dynamics of the spin-ensemble in the interaction frame of the dissipator \eqref{eqn:diss} is given to 2nd order by the TCL master equation~\cite{Bullough1987}:
\bea
\frac{d}{dt} \rho_s(t) &=& \int_{0}^{t-t_0}d\tau \, 
					\tr_c\Big[
						\2 L_I(t) e^{\tau \2 D_c}\2 L_I(t-\tau) \rho_s(t)\otimes\rho_{eq}
					\Big],
					\nonumber\\&&
\label{eqn:me1}
\eea
where $\rho_s(t) = \tr_c[\rho(t)]$ is the reduced state of the spin-ensemble and $\rho_{eq}$ is the equilibrium state of the cavity. Under the condition that $\kappa \gg g\sqrt{N_s}$, the master equation~\eqref{eqn:me1} reduces to
\bea
\frac{d}{dt} \rho_s(t)  &=&\frac{g^2}{4}\int_{0}^{t-t_0}d\tau e^{-\kappa\tau/2}\Big(
					\cos(\Delta \tau) \2 D_s \rho_s(t)
					\nonumber\\
				&&-\sin(\Delta\tau)\2 L_{s}\,\rho(t)
				\Big),
\label{eqn:me2}
\eea
where
\bea
\2 D_s 		&=& 		(1+\overline{n})\,\2 D[J_-]+\overline{n}\, \2D[J_+]		\\
\2 L_{s}\,\rho		&=&		-i [H_{s},\rho]	\\
H_{s}		&=&	 	(1+\overline{n}) \,J_+J_- -\overline{n}\,J_- J_+	
\eea 
are the effective dissipator and Hamiltonian acting on the spin ensemble due to coupling with the cavity.

Under the assumption that $\kappa \gg g\sqrt{N_s}$ we may take the upper limit of the integral in \eqref{eqn:me1} to infinity to obtain the Markovian master equation for the driven spin ensemble:
\be
\frac{d}{dt} \rho_s(t) = \left(\Omega_{s}\,\2 L_{s}+ \frac{\Gamma_s}{2}\, \2 D_s \right)\rho_s(t)
\label{eqn:markovme}
\ee
where 
\bea
\Omega_{s}	= 		-\frac{g^2 \Delta}{\kappa^2 +4\Delta^2}, \quad
\Gamma_s 	= 		\frac{g^2\kappa}{\kappa^2 + 4\Delta^2}.	
\label{eqn:gamma}	
\eea 
Here $\Omega_{s}$ is the frequency of the effective Hamiltonian, and  $\Gamma_s$ is the effective dissipation rate of the spin-system (see~\ref{app:tc-ham}).

%=============================================================
%=============================================================
%		SOLUTION
%=============================================================
%=============================================================
\subsection{Solution to the Markovian master equation}
\label{sec:sol}

We consider the evolution of a spin state which is diagonal in the coupled angular momentum basis, $\rho(t) = \sum_{J}\sum_{m=-J}^J P_{J,m}(t) \rho_{J,m}$. Here the sum over $J$ is summing over subspaces $V_J$, and $P_{J,m}(t)=\bra{J,m}\rho(t)\ket{J,m}$ is the probability of finding the system in the state $\rho_{J,m}=\ketbra{J,m}{J,m}$ at time $t$. In this case the master equation~\eqref{eqn:markovme} reduces to a rate equation for the state populations:
\bea
\frac{d}{dt}P_{J,m}(t)	&=&	
						\Gamma_s\Big(
						A_{J,m+1}P_{J,m+1}(t)
						+ B_{J,m} P_{J,m}(t)
						\nonumber\\
				&&\quad		+ C_{m_J-1} P_{m_J-1}(t) 
						\Big)
\eea
where
\bea
A_{J,m} 	&=& (1+\overline{n})\big[J(J+1)-m(m-1)\big]	\\
C_{J,m}	&=& \overline{n}\big[J(J+1)-m(m+1)\big] 		\\
B_{J,m}	&=& -(A_{J,m}+C_{J,m})
\eea

Defining $\vec{P_J}(t) = (P_{J,-J}(t),\hdots,P_{J,J}(t))$, we obtain the following matrix differential equation for each subspace $V_J$:
\be
\frac{d}{dt} \vec{P_J}(t) = \Gamma_s M_J \vec{P_J}(t),
\label{eqn:rate}
\ee
where $M_J$ is the tridiagonal matrix:
\begin{widetext}
\begin{equation}
M_J = \left(
\begin{matrix}
B_{J,-J}	& A_{J,-J+1} 	& 0 & 0 & 0 & \hdots	&& 0  	\\
C_{J,-J}	& B_{J,-J+1} 	& A_{J,-J+2} & 0 & 0 & \hdots	&& 0  	\\
0 		& C_{J,-J+1} 	& B_{J,-J+2} & A_{J,-J+3} 	& 0 & \hdots 	&& 0  	\\
      \vdots	&			&			&	\ddots	&		&&		& \vdots	\\
     0 	& 	&\hdots	& 0 & C_{J,J-3} 	& B_{J,J-2} & A_{J,J-1}	& 0	\\
     0 	& 	&\hdots	& 0 & 0 		& C_{J,J-2} & B_{J,J-1}	& A_{J,J}	\\
     0 	& 	&\hdots	&0 & 0			& 0		& C_{J,J-1}	& B_{J,J}			
   \end{matrix}
\right)
\label{eqn:ratemat}
\end{equation}
\end{widetext}

For a given state specified by initial populations $\vec P_J(0)$, Eqn~\eqref{eqn:rate} has the solution
\be
\vec P_J(t) = \exp\left(t \,\Gamma_s\,M_J\right)\vec P_J(0).
\label{eqn:rate2}
\ee

%=============================================================
%		EQUILIBRIUM STATES
%=============================================================

The equilibrium state of each subspace $V_J$ of the driven spin-ensemble satisfies $M_J\cdot \vec{P_J}(\infty)=0$, and is given by $\rho_{J,eq}=\sum_{m=-J}^J P_{J,m}(\infty)\rho_{J,m}$, where
\be
P_{J,m}(\infty) = \frac{\overline{n}^{J+m} (1+\overline{n})^{J-m}}{(1+\overline{n})^{2J+1}-\overline{n}^{2J+1}}.
\ee

The total spin expectation value for the equilibrium state of each subspace of the spin-ensemble is
\be
\langle J_x \rangle_{eq} = -J +\overline{n}-\frac{(2J+1)\overline{n}^{2J+1}}{(1+\overline{n})^{2J+1}-\overline{n}^{2J+1}}.
\ee
If we consider the totally symmetric Dicke subspace in the limit of $N_s\gg \overline{n}$, we have that the ground state population at equilibrium is given by $P_{N_s/2,-N_s/2}\approx1/(1+\overline{n})$ and the final expectation value is approximately $\langle J_x \rangle_{eq}  \approx -N_s/2+\overline{n}$. Thus, the final spin polarization in the Dicke subspace will be roughly equivalent to the thermal cavity polarization.

We note that if the detuning $\delta\omega$ were negative, matching $\Omega_R=\delta\omega$ would result in the $H_{+\Omega_R}$ term being dominant, leading to a master equation~\eqref{eqn:markovme} with the operators $J_-$ and $J_+$ interchanged, the dynamics of which would drive the spin ensemble towards the $\langle J_x\rangle=J$ state. Thus, the detuning must be larger than the cavity linewidth to prevent competition between the $H_{-\Omega_R}$ and $H_{+\Omega_R}$ terms, which would drive the spin system to a high entropy thermally mixed state.

%=============================================================
%=============================================================
%		SIMULATIONS
%=============================================================
%=============================================================
\section{Simulations}
\label{sec:sims}

The tridiagonal nature of the rate matrix \eqref{eqn:ratemat} allows Eqn~\eqref{eqn:rate2} to be efficiently simulated for large numbers of spins. For simplicity we will consider the cooling of the Dicke subspace in the ideal case where the cavity is cooled to its ground state ($\overline{n}=0$), and the spin-ensemble is taken to be maximally mixed in the basis of the spin-$J$ subspace ($P_m(0)=1/(2J+1)$ for $m=-J,\hdots,J$). 

\begin{figure}[htbp]
\begin{center}
\includegraphics[width=0.48\textwidth]{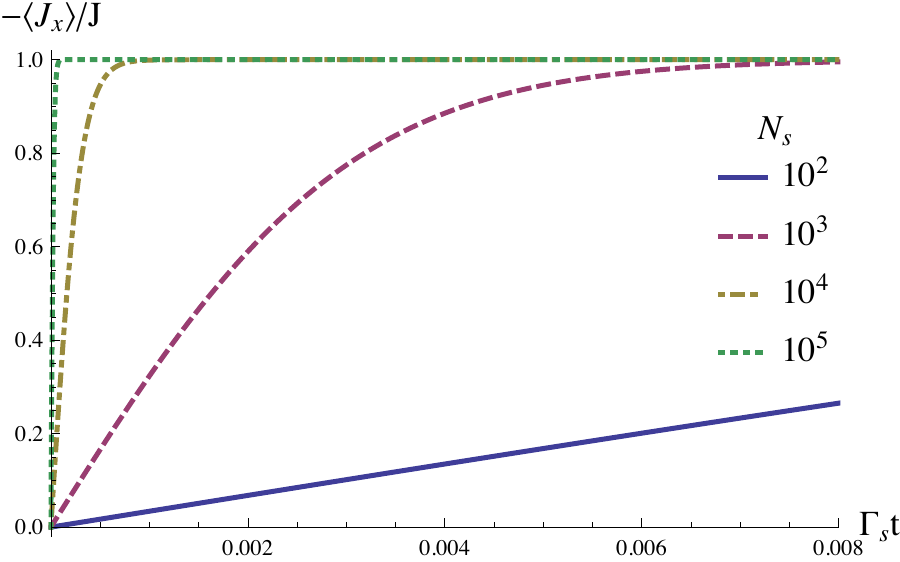}
\caption{Simulated evolution of the normalized expectation value of $-\langle J_x(t)\rangle/J$ for the Dicke subspace of a cavity-cooled spin ensemble. The time axis is scaled by the effective dissipation rate, $\Gamma_s$, for the spin-ensemble given in Eqn~\eqref{eqn:gamma}.}
\label{fig:jx}
\end{center}
\end{figure}

The simulated expectation value of $\langle J_x(t)\rangle$ for the Dicke subspace with total spin $N_s/2$ ranging from $N_s=10^3$ to $10^5$ is shown in Fig.~\ref{fig:jx}, normalized by $-J$ to obtain a maximum value of 1. At a value of $-\langle J_x(t)\rangle/J=1$ the Dicke subspace of the spin ensemble is completely polarized to the $J_x$ ground eigenstate $\ket{J,-J}$.

% FIGURE JX

%=============================================================
%		T_1 FITTING
%=============================================================
%\subsection{Effective $T_1$ time of angular momentum subspaces}

The expectation value $\langle J_x(t)\rangle$ may be fitted to an exponential to derive an effective cooling time-constant, $T_{1,\mbox{\scriptsize{eff}}}$, analogous to the thermal spin-lattice relaxation time, $T_1$. A fit to a model given by
\be
-\frac{\langle J_x(t)\rangle}{J}	 = 1-\exp\left(-\frac{t}{T_{1,\mbox{\scriptsize{eff}}}}\right)
\ee
yields the parameters $T_{1,\mbox{\scriptsize{eff}}}= \lambda(2J)^\gamma/\Gamma_s$ with $\lambda=2.0406$ and  $\gamma=-0.9981$. An approximate expression for the cooling time-constant for the spin subspace $V_J$ as a function of $J$ is then
\be
T_{1,\mbox{\scriptsize{eff}}}(J) \approx \frac{1}{\Gamma_s J} = \frac{\kappa^2+4\Delta^2}{g^2\kappa \,J},
\label{eqn:t1}
\ee
showing that the cooling efficiency is maximized when the Rabi drive strength is matched to the cavity detuning $(\Delta=0)$. In this case the cooling rate and time-constant simplify to $\Gamma_s = g^2/\kappa$ and $T_{1,\mbox{\scriptsize{eff}}}=\kappa/g^2 J$, respectively.

In the case where the cavity is thermally occupied, the final spin polarization is roughly equal to the thermal cavity polarization, and for cavity temperatures corresponding to $\overline{n} < \sqrt{2J}$ the effective cooling constant $T_{1,\mbox{\scriptsize{eff}}}$ is approximately equal to the zero temperature value (see~\ref{app:temp})\cite{supp_mat}.

To achieve this result experimentally, one must choose parameters that adhere to the two RWA's used to isolate the spin-cavity exchange term of Eqn~\eqref{eqn:hint}. Under the condition that $\delta\omega\approx\Omega_R$, this requires that $g\sqrt{N_s}~\ll~\kappa~\ll~\Omega_R,\delta\omega~\ll~\omega_c,\omega_s$ (see~\ref{app:tc-ham})\cite{supp_mat}. For example, assuming an implementation using X-band pulsed electron spin resonance (ESR) ($\omega_c/2\pi\approx\omega_s/2\pi=10$~GHz), with samples that typically contain from roughly $N_s = 10^6$ spins to $N_s = 10^{17}$ spins \cite{Bachar:2012,Eaton2010}, experimentally reasonable values are $\Omega_R/2\pi = 100$~MHz, $Q=10^4$ ($\kappa/2\pi=1$~MHz)\cite{Borneman:2012b,Benningshof:2012a,Malissa:2013a}, and $g/2\pi=1$~Hz \cite{Benningshof:2012a}. 

For these parameters, the range of validity of the Markovian master equation is $N_s\ll \kappa^2/g^2 = 10^{12}$ and the Dicke subspace of an ensemble containing roughly $10^{11}$ electron spins may be polarized with an effective $T_1$ of 3.18~$\mu$s. This polarization time is significantly shorter than the thermal $T_1$ for low-temperature spin ensembles, which normally range from seconds to days \cite{AbragamBook}.

%=============================================================
%=============================================================
%		CONCLUSION
%=============================================================
%=============================================================
\section{Conclusion}
\label{sec:conc}

Several assumptions were made in the presented theoretical model for cavity cooling of a spin ensemble. Firstly, we have assumed that the spin ensemble is magnetically dilute such that no coupling exists between spins. Any spin-spin interaction that breaks the global SU(2) symmetry of the TC Hamiltonian will connect the spin-$J$ subspaces in the coupled angular momentum decomposition of the state space. Such an interaction may be used as an additional resource that should permit complete polarization of the full ensemble Hilbert space. Secondly, we have neglected the effects of thermal relaxation of the spin system. As the cooling effect of the cavity on the spin system relies on a coherent spin-cavity information exchange, the relaxation time of the spin system in the frame of the Rabi drive -- commonly referred to as $T_{1,\rho}$ -- must be significantly longer than the inverse cavity dissipation rate $1/\kappa$. Thirdly, we have assumed that the spin-cavity coupling and Rabi drive are spatially homogeneous across the spin ensemble. Inhomogeneities may be compensated for by numerically optimizing a control pulse that implements an effective spin-locking Rabi drive of constant strength over a range of spin-cavity coupling and control field amplitudes \cite{Borneman:2010a}.

Finally, the derivation of the Markovian master equation~\eqref{eqn:markovme} assumes that no correlations between the cavity and spin system accrue during the cooling process, such that there is no back action of the cavity dynamics on the spin system. This condition is enforced when the cavity dissipation rate, $\kappa$, exceeds the rate of coherent spin-cavity exchange in the lowest excitation manifold by at least an order of magnitude -- i.e. $\kappa \geq 10 g \sqrt{N_s}$ (see~\ref{app:markovian})\cite{supp_mat}. In this Markovian limit, the rate at which spin photons are added to the cavity is significantly less than the rate at which thermal photons are added, meaning the cooling power of the fridge necessary to maintain the thermal cavity temperature is sufficient to dissipate the spin photons without raising the average occupation number of the cavity. From eqn.~\eqref{eqn:t1} we see that, in principle, the cooling efficiency could be improved by adding more spins to make $\kappa$ closer to $g\sqrt{N_s}$, but in this regime the cooling power of the fridge is no longer sufficient to prevent back action from the cavity and strong non-Markovian effects significantly lower the cooling rate. 

%=============================================================
%=============================================================
%		ACKNOWLEDGEMENTS
%=============================================================
%=============================================================
\section*{Acknowledgements}
We thank Holger Haas, Daniel Puzzuoli, Ian Hincks, Christoper Granade, Hamid Mohebbi, and Olaf Benningshof for useful discussions. This work was supported by the Canadian Excellence Research Chairs (CERC) Program and the Canadian Institute for Advanced Research (CIFAR).

%=============================================================
%=============================================================
%		REFERENCES
%=============================================================
%=============================================================
%\bibliography{cavity-cooling-ref}

\begin{thebibliography}{46}
\providecommand{\natexlab}[1]{#1}
\providecommand{\url}[1]{\texttt{#1}}
\expandafter\ifx\csname urlstyle\endcsname\relax
  \providecommand{\doi}[1]{doi: #1}\else
  \providecommand{\doi}{doi: \begingroup \urlstyle{rm}\Url}\fi

\bibitem{Terhal:2013a}
B.M. Terhal. \emph{arXiv:1302.3428v1} (2013).

\bibitem{AbragamBook}
A.~Abragam. \emph{The Principles of Nuclear Magnetism}. Oxford University Press, 1961.

\bibitem{Ramanathan:2008a}
C.~Ramanathan, Appl. Magn. Reson. \textbf{34}, 409 (2008).

\bibitem{Ryan:2008a}
C.A. Ryan, O.~Moussa, J.~Baugh, and R.~Laflamme, Phys. Rev. Lett. \textbf{100}, 140501 (2008).

\bibitem{Weber:2010a}
J.R. Weber, W.F. Koehl, J.B. Varley, A.~Janotti, B.B. Buckley, C.G.~Van
  de~Walle, and D.D. Awschalom, PNAS \textbf{107}, 8513 (2010).

\bibitem{Wineland:1979a}
D.J. Wineland and W.M. Itano, Phys. Rev. A \textbf{20}, 1521 (1979).

\bibitem{Monroe:1995a}
C.~Monroe, D.M. Meekhof, B.E. King, S.R. Jefferts, W.M. Itano, D.J. Wineland,
  and P.~Gould, Phys. Rev. Lett. \textbf{75}, 4011 (1995).

\bibitem{Vuletic:2000a}
V.~Vuletic and S.~Chu, Phys. Rev. Lett. \textbf{84}, 3787 (2000).

\bibitem{Valenzuela:2006a}
S.O. Valenzuela, W.D. Oliver, D.M. Berns, K.K. Berggren, L.S. Levitov, and T.P.
  Orlando, Science \textbf{314}, 1589 (2006).

\bibitem{Wallquist:2008a}
M.~Wallquist, P.~Rabl, M.D. Lukin, and P.~Zoller, New J. Phys. \textbf{10}, 063005 (2008).

\bibitem{Hauss:2008a}
J.~Hauss, A.~Fedorov, A.~Hutter, A.~Shnirman, and G.~Schon, Phys. Rev. Lett. \textbf{100} 037003 (2008).

\bibitem{Murch:2012a}
K.W. Murch, U.~Vool, D.~Zhou, S.J. Weber, S.M. Girvin, and I.~Siddiqi, Phys. Rev. Lett. \textbf{109}, 183602 (2012).

\bibitem{Geerlings:2013a}
K.~Geerlings, Z.~Leghtas, I.M. Pop, S.~Shankar, L.~Frunzio, R.J. Schoelkopf,
  M.~Mirrahimi, and M.H. Devoret, Phys. Rev. Lett. \textbf{110}, 120501 (2013).

\bibitem{Jeener:2002a}
J.~Jeener and F.~Henin, J. Chem. Phys. \textbf{116}, 8036 (2002).

\bibitem{Brahms:2010a}
N.~Brahms and D.~M. Stamper-Kurn, Phys. Rev. A \textbf{82}, 041804(R) (2010).

\bibitem{Hoult:2001a}
D.~I. Hoult and N.~S. Ginsberg, J. Magn. Reson. \textbf{148}, 182 (2001).

\bibitem{Engelke:2010a}
F.~Engelke, Conc. Magn. Reson. A \textbf{36}, 266 (2010).

\bibitem{Tropp:2012a}
J.~Tropp, J. Chem. Phys. \textbf{139}, 014105 (2013).

\bibitem{Hahn:1997a}
E.L. Hahn, Conc. Magn. Reson. \textbf{9}, 69 (1997).

\bibitem{Borneman:2012a}
T.W. Borneman, C.E. Granade, and D.G. Cory, Phys. Rev. Lett. \textbf{108}, 140502 (2012).

\bibitem{Cappellaro:2009a}
P.~Cappellaro, L.~Jiang, J.S. Hodges, and M.D. Lukin, Phys. Rev. Lett. \textbf{102}, 210502 (2009).

\bibitem{Morton:2008a}
J.J.L. Morton, A.M. Tyryshkin, R.M. Brown, S.~Shankar, B.W. Lovett, A.~Ardavan,
  T.~Schenkel, E.E. Haller, J.W. Ager, and S.A. Lyon, Nature \textbf{455}, 1085 (2008).

\bibitem{Staudt:2012a}
M.~U. Staudt, I.-C. Hoi, P.~Krantz, M.~Sandberg, M.~Simoen, P.~Bushev,
  N.~Sangouard, M.~Afzelius, V.~S. Shumeiko, G.~Johannson, P.~Delsing, and
  C.~M. Wilson, J. Phys. B \textbf{45}, 124019 (2012).

\bibitem{Chiorescu:2010a}
I.~Chiorescu, N.~Groll, S.~Bertaina, T.~Mori, and S.~Miyashita, Phys. Rev. B \textbf{82}, 024413 (2010).

\bibitem{Schuster:2010a}
D.I. Schuster, A.P. Sears, E.~Ginossar, L.~DiCarlo, L.~Frunzio, J.J.L. Morton,
  H.~Wu, G.A.D. Briggs, B.B. Buckley, D.D. Awschalom, and R.J. Schoelkopf, Phys. Rev. Lett. \textbf{105}, 140501 (2010).

\bibitem{Kubo:2010a}
Y.~Kubo, F.~R. Ong, P.~Bertet, D.~Vion, V.~Jacques, D.~Zheng, A.~Dreau, J.~F.
  Roch, A.~Auffeves, F.~Jelezko, J.~Wrachtrup, M.~F. Barthe, P.~Bergonzo, and
  D.~Esteve, Phys. Rev. Lett. \textbf{105}, 140502 (2010).

\bibitem{Xiang:2012a}
Z.-L. Xiang, S.~Ashhab, J.Q. You, and F.~Nori, Rev. Mod. Phys. \textbf{85}, 623 (2013).

\bibitem{Bonifacio:1970a}
R.~Bonifacio and G.~Preparata, Phys. Rev. A \textbf{2}, 336 (1970).

\bibitem{Tavis1968}
M. Tavis and F.W. Cummings, Phys. Rev. \textbf{170}, 379 (1968).

\bibitem{Tavis1969}
M. Tavis and F.W. Cummings, Phys. Rev. \textbf{188}, 692 (1969).

\bibitem{Sakurai}
J.~J. Sakurai, \emph{Modern Quantum Mechanics}. Addison Wesley, 1993.

\bibitem{Dicke1954}
R.~H. Dicke, Phys. Rev. \textbf{93}, 99 (1954).

\bibitem{Garraway2011}
B.~M. Garraway, Phil. Trans. Roy. Soc. A \textbf{369}, 1137 (2011).

\bibitem{Bloembergen:1958a}
N.~Bloembergen and P.P. Sorokin, Phys. Rev. \textbf{110}, 865 (1958).

\bibitem{Hartmann:1962a}
S.R. Hartmann and E.L. Hahn, Phys. Rev. \textbf{128}, 2042 (1962).

\bibitem{Belthangady:2013}
C.~Belthangady, N.~Bar-Gill, L.M. Pham, K.~Arai, D.~Le Sage, P.~Cappellaro, and
  R.L. Walsworth, Phys. Rev. Lett. \textbf{110}, 157601 (2013).

\bibitem{supp_mat}
See Supplemental Material at [APS Provided URL] for further details.

\bibitem{Yamamotobook}
Y.~Yamamoto and A.~Imamoglu. \emph{Mesoscopic Quantum Optics}, chapter 6.4. John Wiley \& Sons, Inc., 1999.

\bibitem{Purcell1946}
E.~M. Purcell. In \emph{Proc. Amer. Phys. Soc.} (1946).

\bibitem{Mollow:1969a}
B.R. Mollow, Phys. Rev. \textbf{188}, 1969 (1969).

\bibitem{Breuer2002}
H.-P. Breuer and F. Petruccione. \emph{The theory of open quantum systems}. Oxford University Press, USA, 2002.

\bibitem{Agarwal1974}
G.~S. Agarwal.\emph{Quantum Optics}. Springer-Verlag, Berlin, 1974.

\bibitem{Bullough1987}
R.K.~Bullough. Hyperfine Int. \textbf{37}, 71 (1987).

\bibitem{Bachar:2012}
G.~Bachar, O.~Suchoi, O.~Shtempluck, A.~Blank, and E.~Buks, Appl. Phys. Lett. \textbf{101}, 022602 (2012).

\bibitem{Eaton2010}
G.R. Eaton, S.S. Eaton, R.W. Quine, D.~Mitchell, V.~Kathirvelu, and R.T. Weber, J. Magn. Reson. \textbf{205}, 109 (2010).

\bibitem{Borneman:2012b}
T.W. Borneman and D.G. Cory, J. Magn. Reson. \textbf{225}, 120 (2012).

\bibitem{Benningshof:2012a}
O.W.B. Benningshof, H.R. Mohebbi, I.A.J. Taminiau, G.X. Miao, and D.G. Cory, J. Magn. Reson. \textbf{230}, 84 (2013).

\bibitem{Malissa:2013a}
H.~Malissa, D.~I. Schuster, A.~M. Tyryshkin, A.~A. Houck, and S.~A. Lyon, Rev. Sci. Instr. \textbf{84}, 025116 (2013).

\bibitem{Borneman:2010a}
T.W. Borneman, M.D. Hurlimann, and D.G. Cory, J. Magn. Reson. \textbf{207}, 220 (2010).

%\bibitem{Boero:2013}
%G.~Boero, G.~Gualco, R.~Lisowski, J.~Anders, D.~Suter, and J.~Brugger, J. Magn. Reson. \textbf{231}, 133 (2013).

\end{thebibliography}

%======================================================================
%======================================================================
%		APPENDIX
%======================================================================
%======================================================================

\appendix
\begin{widetext}

\section{Derivation of Markovian Master Equation}
%=======================================
% DERIVATION
%=======================================
\subsection{System Hamiltonian}
\label{app:ham}

We include here a derivation of the interaction Hamiltonian for the Rabi-driven Tavis-Cummings Hamiltonian. Assuming the control field to be on resonance with the Larmor frequency of the spins, the spin-cavity Hamiltonian is given by $H(t) = H_0 + H_R(t) + H_I$, with
\bea
H_0		&=& \omega_c a^\dagger a + \omega_s J_z	\\
H_R(t)	&=&  2\Omega_R \cos(\omega_s t)J_x		\\
H_I		&=& 2g(a^\dagger+a)J_x,
\eea
where $a^\dagger (a)$ are the creation (annihilation) operators describing the cavity, $\Omega_R$ is the strength of the drive field (Rabi frequency), $\omega_c$ is the resonant frequency of the cavity, $\omega_s$ is the Larmor resonance frequency the spins, and $g$ is the coupling strength of the cavity to a single spin in the ensemble in units of $\hbar=1$. Here we have used the notation that $J_\alpha\equiv \sum_{j=1}^{N_s} \sigma_\alpha^{(j)}/2$ are the total angular momentum spin operators for an ensemble of $N_s$ spins. 

After moving into a rotating frame defined by $H_1=\omega_s (a^\dagger a + J_z)$, the spin-cavity Hamiltonian is transformed to
\bea
\widetilde{H}^{(1)}(t)	&=& e^{i t H_1}H\,e^{-i t H_1}-H_1 	\\
				&\approx& \delta\omega a^\dagger a + \Omega_R J_x + g(a^\dagger J_-+a J_+)
\eea
where  $\delta\omega = \omega_c-\omega_s$. The rotating-wave approximation (RWA) used here is valid when the resonant frequencies of the cavity and spin ensemble, $\omega_c$, $\omega_s$, are larger then the inverse time scale we are interested in. In our case this time scale will be dictated by the dissipation rate for the cavity, $\kappa$, and the Hamiltonian frequencies $g$ and $\Omega_R$. Hence we require $\omega_c, \omega_s \gg \kappa,g,\Omega_R$.

If we now move into the interaction frame of $H_2=\delta\omega a^\dagger a + \Omega_R J_x$, the Hamiltonian transforms to
\bea
H_I(t)	&=& e^{i t H_2}H^{(1)}\,e^{-i t H_2}-H_2 	\\
		&=& g\, e^{i t H_2}(a^\dagger J_-+a J_+)e^{-i t H_2}		\\
		&=& g\, e^{i\delta\omega t}a^\dagger \, 
			\Big(
				e^{i \Omega_R J_x t}J_{-} e^{-i \Omega_R J_x t}
			\Big)
			+h.c.		\\
		&=& g\, e^{i\delta\omega t}a^\dagger\Big[
			J_x  
			-i\Big(
				e^{i \Omega_R J_x t} J_y e^{-i \Omega_R J_x t}
			\Big)
			\Big]
			+h.c.	
\eea

Now we use the Baker-Campbell-Hausdorf expansion with
\bea
\mbox{Ad}_{J_x}^0(J_y)	&\equiv& J_y	\\
\mbox{Ad}_{J_x}^1(J_y)	&\equiv& [J_x,J_y] \\
\mbox{Ad}_{J_x}^n(J_y)	&\equiv& \big[J_x,\mbox{Ad}_{J_x}^{n-1}(J_y)\big]
\eea

It follows that for even $n$ $\mbox{Ad}_{J_x}^{n} (J_y) = J_y$, while for odd $n$ $\mbox{Ad}_{J_x}^{n} (J_y) = i J_z$. Hence we have
\bea
e^{i \Omega_R J_x t} J_y e^{-i \Omega_R J_x t}
	&=&	\sum_{n=0}^\infty \frac{\left(i\Omega_R t\right)^n }{n!}
		\mbox{Ad}_{J_x}^n(J_y)		\\
	&=& \sum_{n=0}^\infty \frac{\left(i\Omega_R t\right)^{2n} }{(2n)!}
		\mbox{Ad}_{J_x}^{2n}(J_y)
		+
		\frac{\left(i\Omega_R t\right)^{2n+1}}{(2n+1)!}
		\mbox{Ad}_{J_x}^{2n+1}(J_y)		\\
	&=& 
		\cos\left(\Omega_R t\right) 
		J_y
		-
		\sin\left(\Omega_R t\right) 
		J_z		\\
	&=& \frac{1}{2}\left[
		\cos\left(\Omega_R t\right) 
		(J_+^{(x)}+ J_-^{(x)})
		+i
		\sin\left(\Omega_R t\right) 
		(J_+^{(x)}- J_-^{(x)})
		\right]		\\
	&=&\frac{1}{2}\left(
		e^{i \Omega_R t} \, J_+^{(x)}
		+e^{-i \Omega_R t} \, J_-^{(x)}
		\right)
\eea
where $J_\pm^{(x)}\equiv J_y\pm i J_z$ are the spin-ladder operators in the $x$-basis. Hence we have
\bea
H_I(t)	
		&=& g\, e^{i\delta\omega t}a^\dagger\Big[
			J_x  
			-\frac{i}{2}
				\left(
					e^{i \Omega_R t} \, J_+^{(x)}
					+e^{-i \Omega_R t} \, J_-^{(x)}
				\right)
			\Big]
			+h.c.	
\eea
which may be broken up in terms of frequency components
\bea
H_I(t) &=& H_{0\Omega_R}(t) + H_{-\Omega_R}(t) + H_{+\Omega_R}(t)\\
H_{0\Omega_R}(t)
	&=&  g\Big(
			e^{-i \delta\omega t} a +e^{i\delta\omega t}a^\dagger
		\Big)\,J_x 
					\label{aeqn:hintfull}	\\
H_{-\Omega_R}(t)	
		&=&\frac{i\,g}{2} \Big(
				e^{-i\Delta_- t} a J_+^{(x)}
				- e^{i\Delta_- t} a^\dagger J_-^{(x)} \Big)
				\\
H_{+\Omega_R}(t)
	&=& \frac{i\,g}{2} \Big(
					e^{-i\Delta_+ t} a\, J_-^{(x)} 
					-e^{i\Delta_+ t} a^\dagger J_+^{(x)} 
				\Big)
\eea
where $\Delta_\pm = \delta\omega\pm \Omega_R$.

%=======================================================
%=======================================================
 % DERIVATION OF MASTER EQUATION
%=======================================================
%=======================================================
\subsection{General Markovian master equation}
\label{app:me-deriv}

To model the cavity-induced cooling of the spin system we use an open quantum system description of the cavity and spin ensemble. The joint spin-cavity dynamics may be modelled using the time-convolutionless (TCL) master equation formalism~\cite{Breuer2002}, allowing the derivation of an effective dissipator acting on the spin ensemble alone.

Consider a system Hamiltonian $H(t)$ composed of two general interaction Hamiltonians with a cavity system $H(t)=\sum_\alpha H_\alpha(t)$ where
\bea
H_\alpha(t) 	&=& A_\alpha(t)^\dagger\,a + A_\alpha(t) \,a ^\dagger
\eea
We define the following superoperators for the Liouvillian $\2 L$ describing the unitary portion of the system evolution, and Lindblad dissipator $\2 D$ describing the non-unitary evolution:
\bea
\2 L[A]\rho 	&=& -i [A,\rho ]	\\
\2 D[A]\,\rho	&=& 2 A\rho A^\dagger - \{A^\dagger A,\rho\}
\eea
The evolution of the spin-cavity system is given by the Lindblad master equation
\be
\frac{d}{dt} \rho(t) = \2 L[H(t)]\rho(t) + \2 D_c\rho(t)
\ee
where $\2 L[H(t)]$ is the superoperator describing evolution under the Hamiltonian $H(t)$, and $\2 D_c$ is a dissipator describing the quality factor of the cavity phenomenologically as a photon amplitude damping channel\cite{Agarwal1974}:
\be
\2 D_c =\frac{\kappa}{2}\Big( (1+\overline n)\,\2 D[a] +\overline{n}\,\2 D[a^\dagger]\Big),
\label{aeqn:diss}
\ee
where $\overline{n} = \tr[a^\dagger a \rho_{eq}]$ characterizes the temperature of the bath, and $\kappa$ is the cavity dissipation rate.

We now move into the interaction frame defined by the dissipator $\mcal D_c$. The interaction superoperators in this frame are given by $\widetilde{\mcal S}(t) = e^{-t \mcal D_c}  \mcal S(t) \,e^{t \mcal D_c} $. For density operators we have that $\widetilde{\rho}(t) = e^{-t \mcal D_c}\rho(t)$. Hence we have that our joint system master equation in the dissipator interaction frame is
\be
\frac{d}{dt} \widetilde{\rho}(t) = \widetilde{\2 L}[H_I(t)]\widetilde{\rho}(t)
\ee

We define a projection operator $\mcal P$ onto the relevant degrees of freedom for our reduced system
\be
\mcal P \rho(t)\equiv \rho_s(t)\otimes\rho_{eq}
\ee
where $\rho_s(t)=\tr_c \left[ \rho(t) \right]$, and $\rho_{eq}$ is the equilibrium state of the cavity under the dissipator $\2 D_c$ ($\2 D_c \rho_{eq}=0$). In the case of weak coupling, the second order TCL master equation is given by~\cite{Breuer2002}
\begin{equation}
\frac{d}{dt}\mcal P \widetilde{\rho}(t) = \int_{0}^{t-t_1} d \tau \,
 \mcal P\,\widetilde{\mcal L}[H(t)]\widetilde{\mcal L}[H(t-\tau)]\mcal P\widetilde{\rho}(t).
\end{equation}

We now explicitly consider the interaction frame of the dissipator. To do this we use the definition of the adjoint channel $\mcal D_c^\dagger$ which satisfies $\tr_c [\mcal D_c^\dagger[A] \,B]=\tr_c[A \mcal \,D_c[B]]$ for all operators $A,B$ on the cavity system. 
The adjoint channel has the following useful properties:
\bea
\mcal D_c^\dagger[\id]			&=&	0, \quad\quad
\mcal D_c^\dagger[a]			= 	-\frac{\kappa}{2} a, 	\quad\quad
\mcal D_c^\dagger[a^\dagger]		=	 -\frac{\kappa}{2} a^\dagger	\\
 e^{t \mcal D_c^\dagger}[\id]		&=&	1, \quad\quad
 e^{t \mcal D_c^\dagger}[a]			= 	e^{-\frac{\kappa}{2}t} a, 	\quad\quad
 e^{t \mcal D_c^\dagger}[a^\dagger]	=	e^{-\frac{\kappa}{2}t} a^\dagger
\eea

Hence we have that
\bea
\mcal P \widetilde{\rho}(t) 
&=& \tr_c [ e^{-t \mcal D_c}\rho(t)]\otimes\rho_{eq}
=  \tr_c [ e^{-t \mcal D^\dagger_c}[\bb I]\, \rho(t)]\otimes\rho_{eq}
= \mcal P \rho(t).
\end{eqnarray}
In addition also $\mcal D_c \mcal P \rho(t)= \tr_c[\rho]\otimes \mcal D_c\rho_{eq}=0$, thus the reduced dynamics of the spin-ensemble in the interaction frame of the dissipator \eqref{eqn:diss} is given to 2nd order by the TCL master equation~\cite{Bullough1987}:
\bea
\frac{d}{dt} \rho_s(t) &=& \int_{0}^{t-t_0}d\tau \, 
					\tr_c\Big[
						\2 L[H(t)] e^{\tau \2 D_c}\2 L[H(t-\tau)] \rho_s(t)\otimes\rho_{eq}
					\Big].
\label{aeqn:me1}
\eea

Now using the properties of the adjoint dissipator we have
\bea
\frac{d}{dt} \rho_s(t) 	&=& \int_{0}^{t-t_0}d\tau \, 
					\tr_c\Big[e^{\tau \2 D^\dagger_c}(
						\2 L[H(t)]) \2 L[H(t-\tau)] \rho_s(t)\otimes\rho_{eq}
					\Big]	\nonumber\\
				&=& \int_{0}^{t-t_0}d\tau \,e^{-\kappa \tau/2} 
					\tr_c\Big[
						\2 L[H(t)]\, \2 L[H(t-\tau)] \rho_s(t)\otimes\rho_{eq}
					\Big]	\nonumber\\
				&=& -\int_{0}^{t-t_0}d\tau \,e^{-\kappa \tau/2} \,
					\tr_c\Big[
						H(t), \Big[ H(t-\tau), \rho_s(t)\otimes\rho_{eq}
					\Big]\Big].	
				\label{aeqn:me-genH}
\eea

Starting with the 2nd order TCL master equation~\eqref{eqn:me-genH}, we now expand this in terms of the component Hamiltonians $H_\alpha(t)$.

Define
\bea
\2 C_{\alpha,\beta}(t,s)	&=& \tr_c \Big[	
							H_\alpha(t), \big[ H_\beta(s),
							\rho_s(t)\otimes\rho_{eq}\big]
							\Big]
\eea

Hence we have
\bea
\frac{d}{dt} \rho_s(t) 	&=& -\sum_{\alpha,\beta}
					\int_{0}^{t-t_0}d\tau \,e^{-\kappa \tau/2} \,						
						\2 C_{\alpha,\beta}(t,t-\tau)
\eea

Using the properties of our cavity equilibrium state 
\bea
	\tr[aa^\dagger \rho_{eq}]&=&\overline{n}+1,  	\quad\quad
	\tr[a^\dagger a\rho_{eq}]=\overline{n},		\quad\quad
	\tr[a^2 \rho_{eq}]=\tr[a^{\dagger 2} \rho_{eq}]=0, 
\eea	
we have only two contributing terms for each $\2 C$. Hence we have that $\2 C_{\alpha,\beta}(t,s)$ is given by
\bea
\2 C_{\alpha,\beta}(t,s)
		&=& \tr_c \Big[	A_\alpha(t)^\dagger a, 
				\big[A_\beta(s)a^\dagger,\rho\otimes\rho_{eq}\big]\Big]
			+\tr_c \Big[	A_\alpha(t)a^\dagger , 
				\big[A_\beta(s)^\dagger a,\rho\otimes\rho_{eq}\big]\Big]		\\
		&=& (\overline{n}+1)
				\Big(
					A_\alpha(t)^\dagger A_\beta(s)\, \rho 
					+ \rho\, A_\beta(s)^\dagger A_\alpha(t)
					- A_\beta(s)\,\rho \,A_\alpha(t)^\dagger 
					- A_\alpha(t) \,\rho\, A_\beta(s)^\dagger
				\Big)
		\nonumber\\
		&&+\overline{n}\,
				\Big(
					A_\alpha(t) A_\beta(s)^\dagger\, \rho 
					+ \rho\, A_\beta(s) A_\alpha(t)^\dagger
					- A_\beta(s)^\dagger \,\rho\, A_\alpha(t) 
					- A_\alpha(t)^\dagger \,\rho\, A_\beta(s)
				\Big)
\eea

Now to calculate the dissipator for these terms we have that in the Markovian limit we take the upper limit of the integral to infinity $\int_{0}^{t-t_0}d\tau \rightarrow \int_{0}^{\infty}d\tau$, and we define the superoperator generators

\bea
\2 G_{\alpha}(t) &=& -\int_{0}^{\infty} d\tau \, 
					e^{-\kappa \tau/2}
					\2 C_{\alpha,\alpha}(t,t-\tau)	\\
\2 G_{\alpha,\beta}(t) &=& -\int_{0}^{\infty} d\tau \, 
						e^{-\kappa \tau/2}
						\Big( 
						\2 C_{\alpha,\beta}(t,t-\tau) 
						+\2 C_{\beta,\alpha}(t,t-\tau)
						\Big)
\eea
Hence our reduced system master equation is given by
\bea
\frac{d}{dt}\rho_s(t) 	&=&\Big(
					\sum_\alpha\2 G_{\alpha}(t)
					+\sum_{\alpha<\beta} \2 G_{\alpha,\beta}(t) 
					\Big)\rho_s(t)
\eea

The $\2G_\alpha(t)$, and $\2G_{\alpha,\beta}(t)$ terms we refer to as the diagonal and cross-terms of the master equation respectively. In general, inclusion of the cross terms will lead to a master equation for the spin system which does not generate a completely positive map, however we may remove them under certain parameter regimes with an appropriate RWA.

Suppose that the time dependence of the operators $A_\alpha(t)$ is such that $A_\alpha(t) = e^{i \omega_\alpha t }A_\alpha$. Hence $A_\alpha(t-\tau)=e^{-i\tau\omega_\alpha}A(t)$, and in this case we have
\bea
\int_{0}^{\infty} d\tau \, e^{-\kappa \tau/2}e^{\pm i \tau\omega_\alpha}
	&=&		\frac{2}{\kappa\mp i 2\omega_\alpha}  
	= 2 \left(
				\frac{\kappa\pm2i\omega_\alpha}{\kappa^2+4\omega_\alpha^2}  
		\right)
	= \gamma_\alpha \pm i\lambda_\alpha
\eea
where
\be
\gamma_\alpha = \frac{2\kappa}{\kappa^2+4\omega_\alpha^2} 
\quad\quad
\lambda_\alpha = \frac{4\omega_\alpha}{\kappa^2+4\omega_\alpha^2}.
\ee

In this case the cross-terms $\2 G_{\alpha,\beta}$ will still have time dependence of $e^{\pm i(\omega_\alpha-\omega_\beta)t }$. Thus, if we have that $|\omega_\alpha-\omega_\beta| \gg \kappa $ for all $\alpha,\beta$, then we may make a RWA and disregard these high frequency terms.

After making this RWA we have
\bea
\2 G_\alpha(t) \rho
		&=& -\int_{0}^{\infty} d\tau\, e^{-\kappa\tau/2}\2 C_{\alpha,\beta}(t,t-\tau)\rho	\\
		&=&-\int_{0}^{\infty} d\tau\, e^{-\kappa\tau/2}
			\Bigg[ 
				(\overline{n}+1)
				\Big(
					e^{-i\omega_\alpha\tau}(A_\alpha^\dagger A_\alpha\, \rho 
					-A_\alpha\,\rho \,A_\alpha^\dagger)
					+ e^{i\omega_\alpha\tau}(\rho\, A_\alpha^\dagger A_\alpha
					-A_\alpha \,\rho\, A_\alpha^\dagger)
				\Big)
		\nonumber\\
		&&+\overline{n}\,
				\Big(
					e^{i\omega_\alpha\tau}(A_\alpha A_\alpha^\dagger\, \rho 
					- A_\alpha^\dagger \,\rho\, A_\alpha )
					+e^{-i\omega_\alpha\tau} (\rho\, A_\alpha A_\alpha^\dagger
					- A_\alpha^\dagger \,\rho\, A_\alpha)
				\Big)
				\Bigg]	\\
		&=&	(\overline{n}+1)
				\Big(
					(\gamma_\alpha-i\lambda_\alpha)
					(A_\alpha^\dagger A_\alpha\, \rho 
					-A_\alpha\,\rho \,A_\alpha^\dagger)
					+ (\gamma_\alpha+i\lambda_\alpha)
					(\rho\, A_\alpha^\dagger A_\alpha
					-A_\alpha \,\rho\, A_\alpha^\dagger)
				\Big)
		\nonumber\\
		&&+\overline{n}\,
				\Big(
					(\gamma_\alpha+i\lambda_\alpha)
					(A_\alpha A_\alpha^\dagger\, \rho 
					- A_\alpha^\dagger \,\rho\, A_\alpha )
					+(\gamma_\alpha-i\lambda_\alpha)
					(\rho\, A_\alpha A_\alpha^\dagger
					- A_\alpha^\dagger \,\rho\, A_\alpha)
				\Big) 	\\
	&=& 
	\gamma_\alpha
		\bigg(
			(\overline{n}+1)		\2 D[A_\alpha]
			+\overline{n}\,D[A_\alpha^\dagger]
		\bigg)
	-\lambda_\alpha\,
		\2L\Big[(\overline{n}+1)A_\alpha^\dagger A_\alpha
			-\overline{n}A_\alpha A_\alpha^\dagger\Big]
\eea

Hence 
\bea
\2 G_{\alpha}(t) = \2 G_\alpha 
	&=& 
	\gamma_\alpha
		\bigg(
			(\overline{n}+1)		\2 D[A_\alpha]
			+\overline{n}\,D[A_\alpha^\dagger]
		\bigg)
	-\lambda_\alpha\,
		\2L\Big[(\overline{n}+1)A_\alpha^\dagger A_\alpha
			-\overline{n}A_\alpha A_\alpha^\dagger\Big]
\eea
since the time dependence of $A_\alpha(t)$ drops out in the dissipator and Liouvilians. 

%=======================================================
%=======================================================
 % DERIVATION OF MASTER EQUATION
%=======================================================
%=======================================================
\subsection{Markovian master equation for the Rabi-driven TC-Hamiltonian}
\label{app:tc-ham}

We now derive the 2nd order TCL master equation for the Rabi-driven TC-Hamiltonian:
\bea
H_I(t) &=& H_{0\Omega_R}(t) + H_{-\Omega_R}(t) + H_{+\Omega_R}(t)\\
H_{0\Omega_R}(t)
	&=&  g\Big(
			e^{-i \delta\omega t} a +e^{i\delta\omega t}a^\dagger
		\Big)\,J_x 
					\label{aeqn:hintfull}	\\
H_{-\Omega_R}(t)	
		&=&\frac{i\,g}{2} \Big(
				e^{-i\Delta_- t} a J_+^{(x)}
				- e^{i\Delta_- t} a^\dagger J_-^{(x)} \Big)
				\\
H_{+\Omega_R}(t)
	&=& \frac{i\,g}{2} \Big(
					e^{-i\Delta_+ t} a\, J_-^{(x)} 
					-e^{i\Delta_+ t} a^\dagger J_+^{(x)} 
				\Big)
\eea
where $\Delta_\pm = \delta\omega\pm \Omega_R$. In this case the cross terms for the 2nd order TCL master equation will be of frequencies $\Omega_R$ and  $2\Omega_R$, hence our RWA is valid in the regime where the Rabi drive strength is much stronger than the dissipation rate ($\Omega_R\gg\kappa$). In this case we have have three contributions to the master equation for the spin-ensemble:

\bea
\frac{d}{dt} \rho_s(t) 
			&=&\Big(	\2 G_{H_{0\Omega_R}}
					+\2 G_{H_{-\Omega_R}}
					+\2 G_{H_{+\Omega_R}}\Big)\rho_s(t)
\eea
where
\bea
\2 G_{0,\pm\Omega_R}
			&=&
				-\Omega_{0,\pm} \2 L_{0,\pm}
				+\frac{\Gamma_{0,\pm}}{2} \2 D_{0,\pm}
\eea
with
\[
\begin{array}{ll}
\Gamma_{0}	=	\frac{4g^2\kappa}{\kappa^2 + 4\delta\omega^2}
&\quad\quad
\Omega_{0}	=	\frac{4g^2\delta\omega}{\kappa^2 + 4\delta\omega^2}\\
\Gamma_{-}	=	\frac{g^2\kappa}{\kappa^2 + 4\Delta_-^2}
&\quad\quad
\Omega_{-}	=	\frac{g^2\Delta_-}{\kappa^2 + 4\Delta_-^2}\\
\Gamma_{+}	=	\frac{g^2\kappa}{\kappa^2 + 4\Delta_+^2}	
&\quad\quad
\Omega_{+}	=	\frac{g^2\Delta_+}{\kappa^2 + 4\Delta_+^2}\\	
\2 D_{0}		=	(2\overline{n}+1) \2 D[J_x]
&\quad\quad	
\2 L_{0}		=	 \2 L[J_x^2]	\\
\2 D_{-}		=	(\overline{n}+1) \2 D[J_-^{(x)}]+ \overline{n}\, \2 D[J_+^{(x)}]	
&\quad\quad	
\2 L_{-}		=	\2 L[(\overline{n}+1)\,J_+^{(x)} J_-^{(x)} -\overline{n}J_-^{(x)} J_+^{(x)}\big] \\
\2 D_{+}		=	(\overline{n}+1) \2 D[J_+^{(x)}]+ \overline{n}\, \2 D[J_-^{(x)}]	
&\quad\quad	
\2 L_{+}		=	 \2 L[(\overline{n}+1\,J_-^{(x)} J_+^{(x)} -\overline{n}J_+^{(x)} J_-^{(x)}\big]\\	
\end{array}
\]

To achieve cavity cooling to the ground state we require that the $\2G_{-\Omega_R}$ term be dominant, which implies $\Gamma_- \gg\Gamma_+,\Gamma_0$. If we assume that our Rabi drive and cavity detuning are matched, $\Omega_R\approx \delta\omega$, then in the regime where our RWA is valid ($\Omega_R\gg\kappa$), we have
\be
\frac{\Gamma_+}{\Gamma_-} 
	= \frac{\kappa^2+4\Delta_-^2}{\kappa^2+4\Delta_+^2}
	\approx	\frac{\kappa^2}{\kappa^2+16\Omega_R^2}
	\ll 1
\ee
and
\be
\frac{\Gamma_0}{\Gamma_-} 
	= 4\frac{\kappa^2+4\Delta_-^2}{\kappa^2+4\delta\omega^2}
	\approx	4\frac{\kappa^2}{\kappa^2+4\Omega_R^2}
	\ll 1
\ee
and so $\2 G_{-\Omega_R}$ will be the dominant dissipative term. We also have that $\Omega_0 \approx g^2/\delta\omega \ll 1, \Omega_+ \approx g^2/\Delta_+ \ll 1$ since $\delta\omega,\Delta_+\gg g$. Thus, we arrive at the Markovian master equation used in the main text:
\bea
\frac{d}{dt} \rho_s(t) 
			&=&\Big(
				-\Omega_- \2L_- + \frac{\Gamma_-}{2}\2 D_-
				\Big)
				\rho_s(t).
\eea

%=======================================
% VALIDITY OF MARKOV
%=======================================
\subsection{Validity of Markov Approximation}
\label{app:markovian}
The validity of the derived cooling rates depends on the validity of the Markov approximation used to derive the master equation. As stated in the main text, enforcing the condition that the cavity dissipation rate, $\kappa$, exceeds the rate of coherent exchange between the spins and cavity implies that there will be no back action of the cavity dynamics on the spin system. This condition is given by $\kappa\gg g\sqrt{N_s}$, where $N_s$ is the number of spins in the ensemble, and ensures that any photons transferred to the cavity are dissipated in the cavity before they have any back action on the spin system. 

% FIGURE Tavis-Cummings Energy Levels
\begin{figure}[htbp]
\begin{center}
\includegraphics[width=0.85\textwidth]{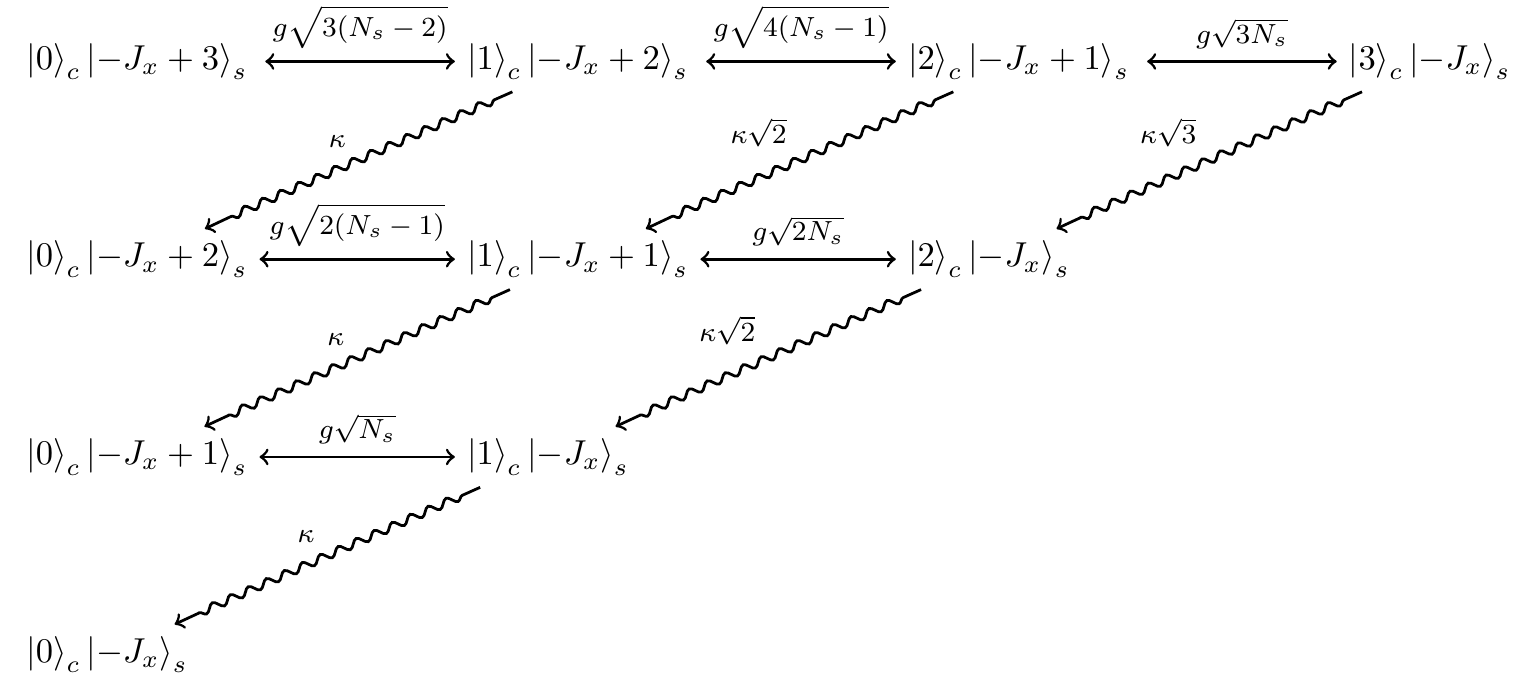}
\caption{Energy level diagram of the joint spin-cavity system with coherent transitions denoted by a solid line and cavity dissipation rates denoted by a curved line. States are labelled as $\ket{n}_c\ket{-J_x + m}_s$, where $m$ is the number of spin excitations and $n$ is the number of cavity excitations. For the cooling dynamics to appear Markovian states of high cavity excitation number should not be significantly populated on a coarse-grained time scale.}
\label{fig:tc-energy-levels}
\end{center}
\end{figure}

More concretely, from the spin-cavity energy level diagram shown in Fig. \ref{fig:tc-energy-levels}, the rate of transfer between states $\ket{n}_c\ket{-J_x+m}_s$ and $\ket{n+1}_c\ket{-J_x+m-1}_s$ is given by $g\sqrt{m(2 J_x+1-m)}\sqrt{n+1}$. At the same time, the cavity dissipator of strength $\kappa\sqrt{n+1}$ is acting to drive the spin-cavity system to the state $\ket{n}_c\ket{-J_x + m -1}_s$. To satisfy the Markov condition, we require the cooling dynamics to always drive the spin-cavity system toward states of low excitation number (bottom left of diagram), without significantly populating states of high excitation number (top right of diagram). This will occur if the maximum rates for coherent transfer and cavity dissipation obey the following relationship:
\begin{equation}
\kappa\sqrt{n+1} \gg g\sqrt{m(2 J_x+1-m)}\sqrt{n+1} \Longleftrightarrow \kappa\gg g\sqrt{m(N_s+1-m)}.
\end{equation}
This transfer rate is greatest for a maximally excited spin system with $m = 2J_x = N_s$. In this case we recover our condition that
\begin{equation}
\kappa \gg g\sqrt{N_s}.
\end{equation}

% FIGURE Jx vs Kappa
\begin{figure*}[htbp]
\begin{center}
\includegraphics[width=\textwidth]{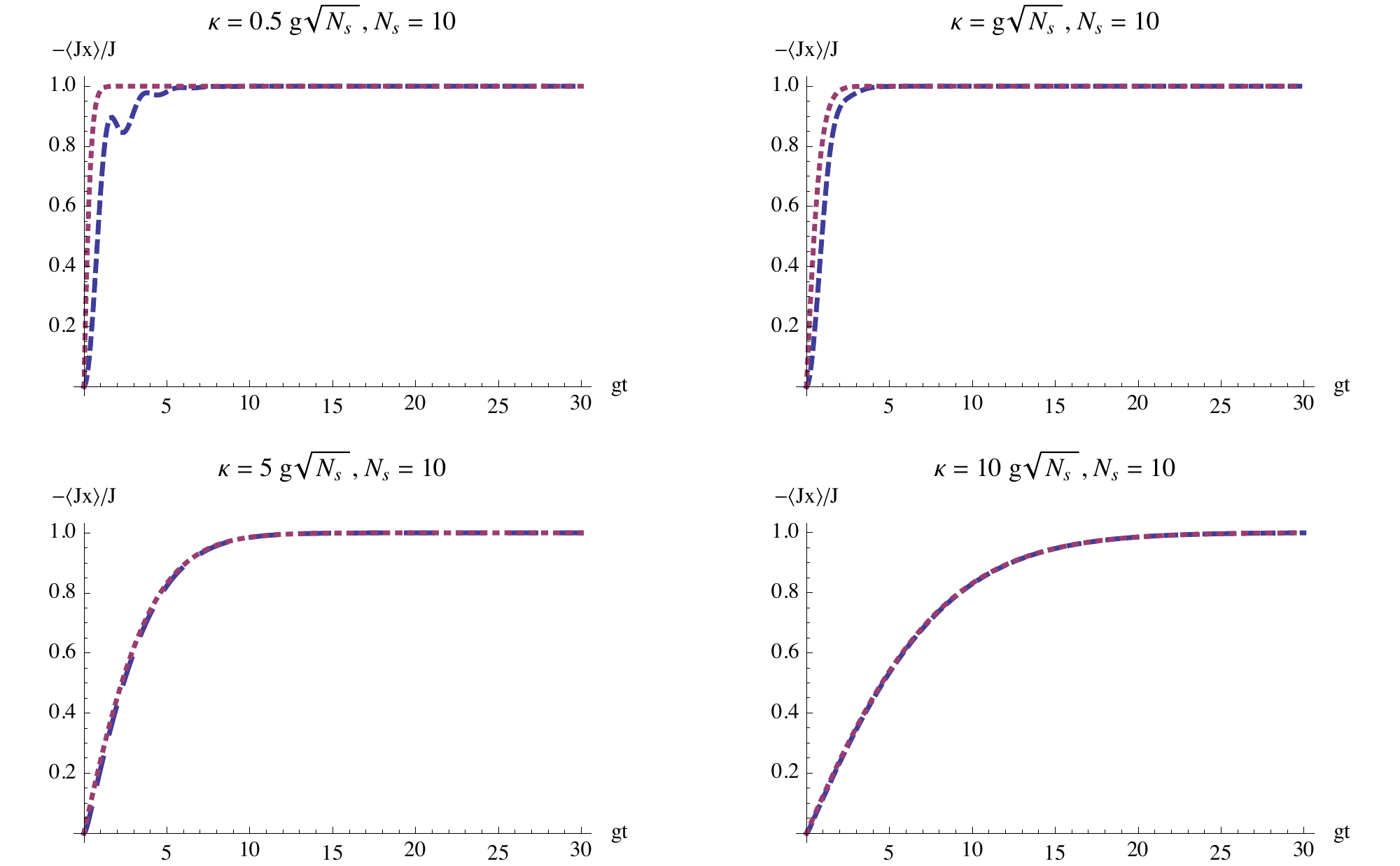}
\caption{Comparison of the cooling dynamics of the Dicke subspace by the Markovian master equation (pink dotted curve) and a spin-ensemble cavity simulation with RWA (blue dashed curve). The normalized expectation value of $-\langle J_x(t)\rangle/J$ is plotted for $N_s=10$ spins with $\kappa= 0.5, 1, 5, 10 g\sqrt{N_s}$, a cavity temperature $T=0K$, and Rabi-drive matched to the detuning $\Delta=\delta\omega-\Omega_r = 0$. When $\kappa \geq 10 g \sqrt{N_s}$ the Markovian master equation calculation agrees very well with the full simulation. Also, as predicted by eqn. (29), the cooling rate increases for larger $\kappa$, until the point where non-Markovian effects take over.}
\label{fig:fullvsmarkov}
\end{center}
\end{figure*}

To numerically investigate where the Markovian approximation breaks down we simulated the evolution for the Dicke subspace of the spin cavity system for $N_s=10$ spins at zero cavity temperature ($\overline{n}=0$) with $\delta\omega=\Omega_R$ in the parameter regime where the RWA is valid and compared it to the derived master equation for the spin ensemble. 
For the full evolution we consider
\bea
\rho(t) = \exp\left[ t\left(\2 L[H_I]+\frac\kappa2 \2D[a]\right)\right] \rho(0)
\eea
where $H_I=\frac{i g}{2}\left( a J_+^{(x)}-a^\dagger J_-^{(x)}\right)$, $\rho(0)=\rho_s(0)\otimes\rho_{eq}$, $\rho_s(0)$ is the maximally mixed state, and $\rho_{eq}=\ketbra{0}{0}$ is the ground state of the cavity. For simulation we truncate the cavity dimension to be $N_s+1$, the dimension of the Dicke subspace of the spin-ensemble.

For the Markovian master equation we consider the rate equation derived in the main paper with $\overline{n}=0$:
\be
\frac{d}{dt} \vec{P}(t) = \frac{g^2}{\kappa} M \vec{P}(t),
\label{aeqn:rate}
\ee
where $M$ is the tridiagonal matrix
\be
M = \left(
   \begin{matrix}
      -A_{-J}& 	A_{-J+1} 	& 	0 		& 0 			& 0		& \hdots	& 0  	\\
      0		&	-A_{-J+1} 	& 	A_{-J+2} 	& 0 			& 0		& \hdots	& 0  	\\
      0		&	0		&-A_{-J+2} 	& 	A_{-J+3} 	& 0 		& \hdots 	& 0  	\\
      \vdots	&			&			&	\ddots	&		&		& \vdots	\\
      0 	& 			&	\hdots	&			& 0		& 0		& -A_{J}			
   \end{matrix}
\right)
\label{aeqn:ratemat}
\ee
with $A_m = (1+\overline{n})\big[J(J+1)-m(m-1)\big]$, $\vec{P}(t) = (P_{-J}(t),\hdots,P_{J}(t))$, and $P_m(t)=\bra{J,m}\rho(t)\ket{J,m}$ is the probability of finding the system in the Dicke state $\rho_m=\ketbra{J,m}{J,m}$ at time $t$.

As shown in Fig. \ref{fig:fullvsmarkov}, when $\kappa = 0.5g\sqrt{N_s}$ a full non-Markovian simulation of the cooling procedure yields dynamics that are richer than predicted by the Markovian model used in the main text. In particular, coherent transfer of spin photons deposited in the cavity back to the spin system are seen as oscillations in the expectation value of $J_x$. These memory effects reduce the cooling efficiency such that the cooling rate is initially fast when the cavity occupation is low, then slows down significantly as higher excitations of the cavity are transferred back to the spin system. As $\kappa$ becomes larger the oscillations are damped out, but the Markovian master equation still does not fully agree with the full non-Markovian simulation until we have $\kappa\approx10g\sqrt{N_s}$. When $\kappa = 10g\sqrt{N_s}$, the oscillations are critically damped and the Markovian master equation captures the full cooling dynamics. Thus, if the cavity dissipation rate, $\kappa$, exceeds the rate of coherent spin-cavity exchange in the single excitation manifold by at least an order of magnitude --- i.e. $\kappa \geq 10 g \sqrt{N_s}$ --- then the Markovian master equation is valid.

%=======================================
% THERMALLY EXCITED CAVITY
%=======================================
\section{Thermally excited cavity}
\label{app:temp}

In the Markov master equation simulations used to derive the effective cooling constant, $T_{1,\mbox{\scriptsize{eff}}}$, Eqn. (29), as a function of ensemble size, $N_s$, we assumed that the cavity was initially cooled to its ground state. The fit of $T_{1,\mbox{\scriptsize{eff}}}$ for cooling simulations of the Dicke subspace of a spin ensemble with non-zero average cavity occupation number $\overline{n} =\left(e^{\omega_c/k_B T_c}-1\right)^{-1}$, is show in Fig~\ref{fig:t1vst}. Here $\omega_c$ and $T_c$ are the resonant frequency and equilibrium temperature of the cavity. The effective cooling time was found to obey
\be
T_{1,\mbox{\scriptsize{eff}}} \approx 
\left\{
\begin{array}{l}
\frac{2}{N_s} 			\quad\mbox{for } \overline{n}< \sqrt{N_s} 	\\
\frac{1}{2\overline{n}} 	\quad\mbox{for } \overline{n} \gtrsim N_s 	\\
\end{array}
\right.
\ee

% FIGURE T1 Vs Temp
\begin{figure}[htbp]
\begin{center}
\includegraphics[width=0.47\textwidth]{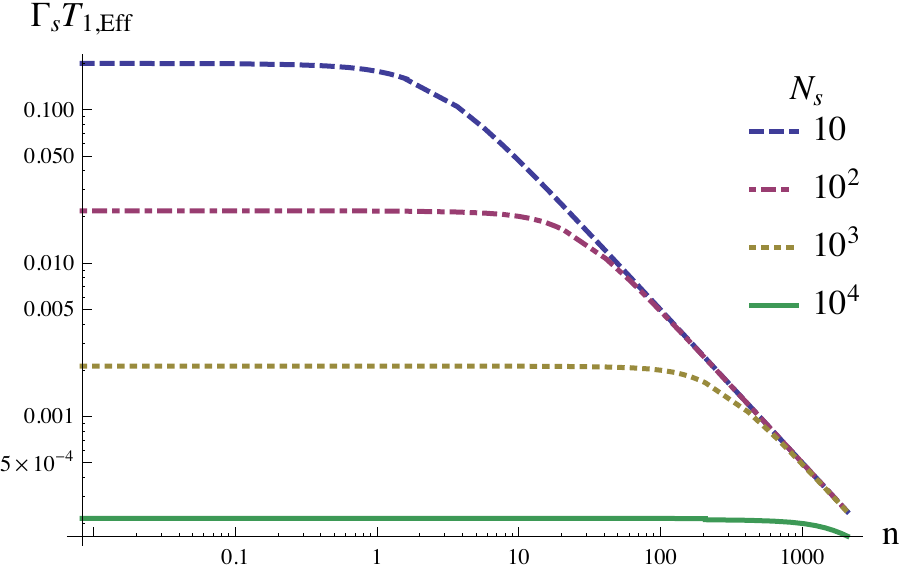}
\caption{Effective cooling time-constant, $T_{1,\mbox{\scriptsize{eff}}}$, of the Dicke subspace of a spin ensemble as a function of the equilibrium excitation number of the cavity, $\overline{n}$, for $N_s=10, 10^2, 10^3, 10^4$ spins in the ensemble.}
\label{fig:t1vst}
\end{center}
\end{figure}

This effect appears to originate from the fact that the final spin system polarization will be equal to the cavity polarization. Thus, cooling to a spin temperature that is not fully polarized requires removing fewer photons from the spin system. Given that the spin dissipation rate is independent of the cavity temperature, it takes less time to drive the spins to a state that is not fully polarized. 
An example of the simulated normalized spin expectation value $-\langle J_x\rangle/J$ as a function of temperature is shown in Fig~\ref{fig:jxvst}. Here we are considering a cavity with resonant frequency of $\omega_c/2\pi =$ 10 GHz.
 
% FIGURE T1 Vs Temp
\begin{figure}[htbp]
\begin{center}
\includegraphics[width=0.45\textwidth]{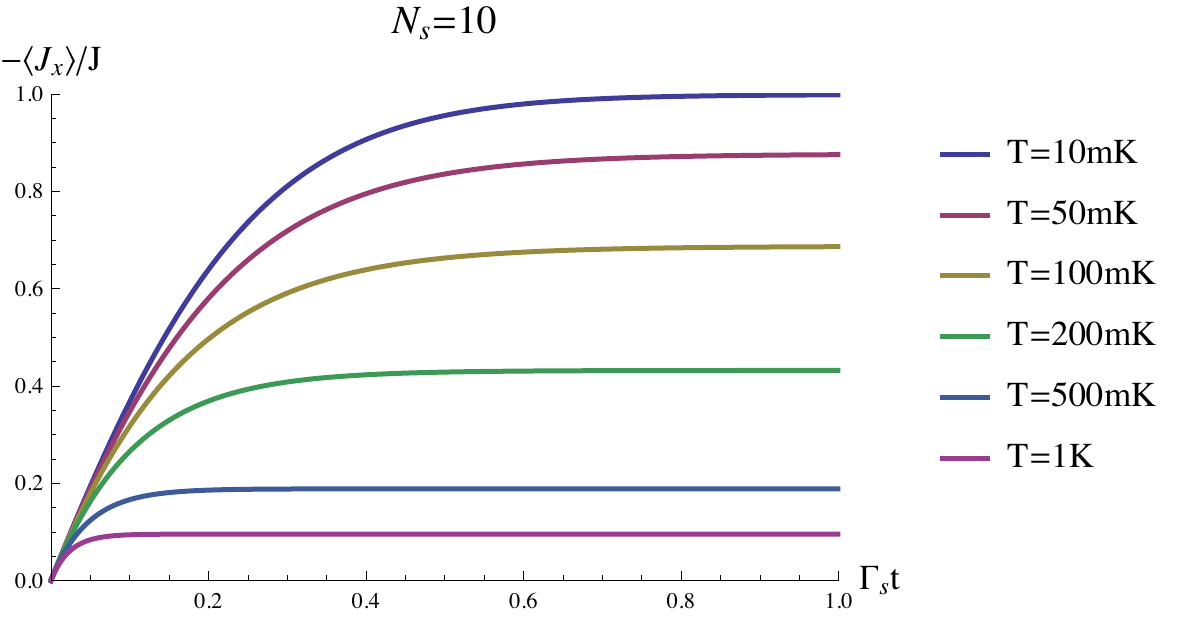}\quad
\includegraphics[width=0.45\textwidth]{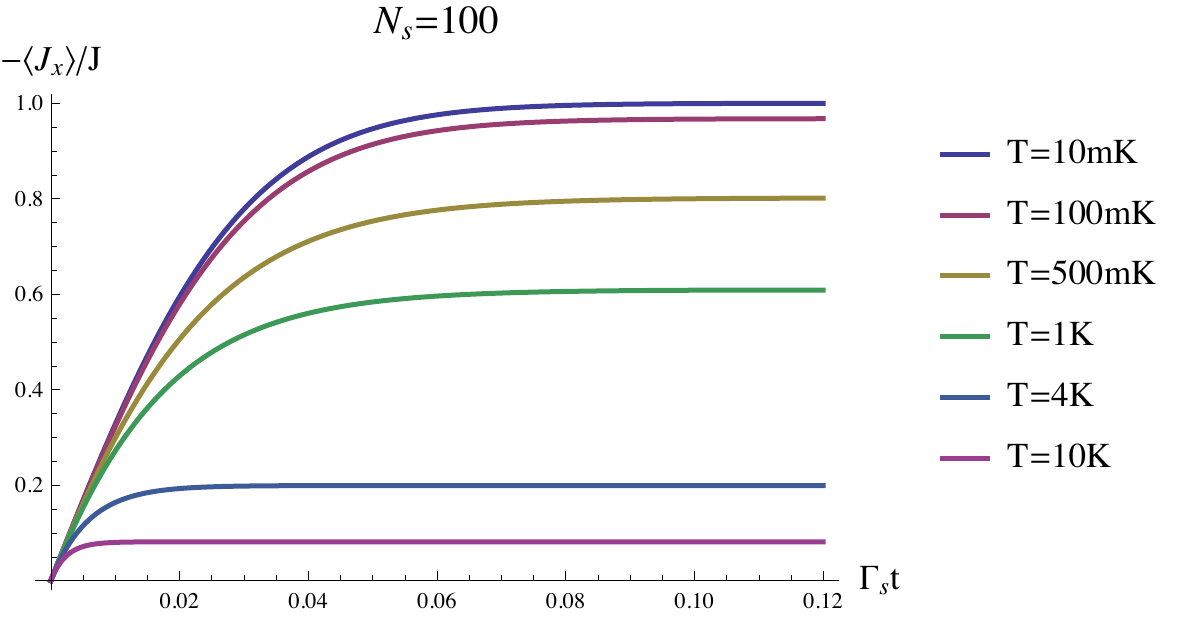}
\includegraphics[width=0.45\textwidth]{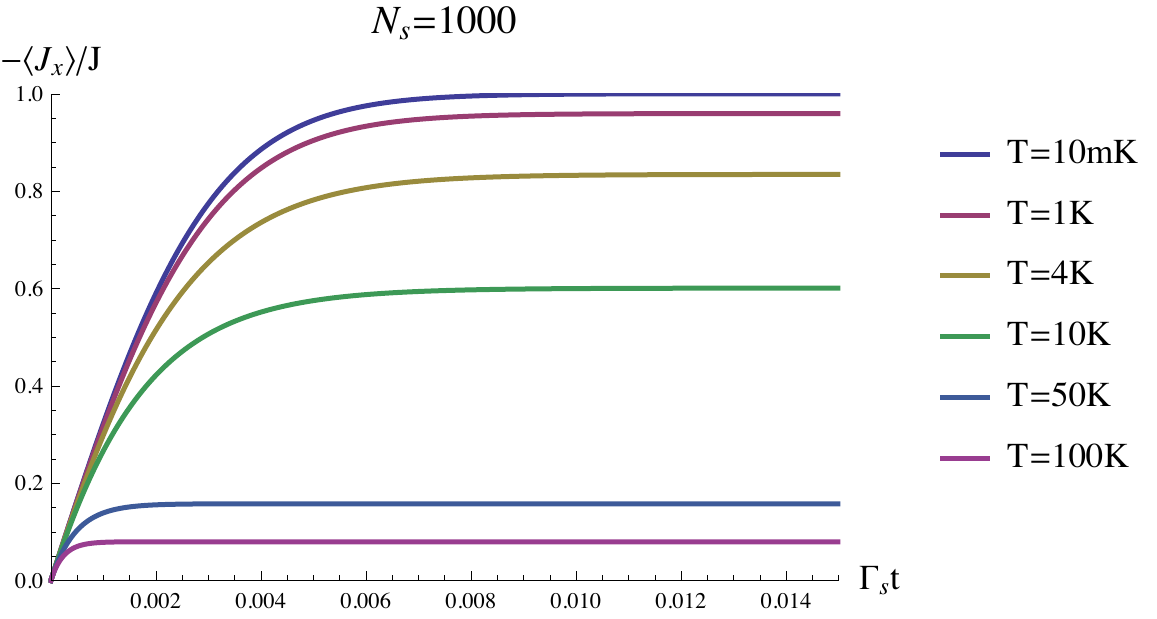}\quad
\includegraphics[width=0.45\textwidth]{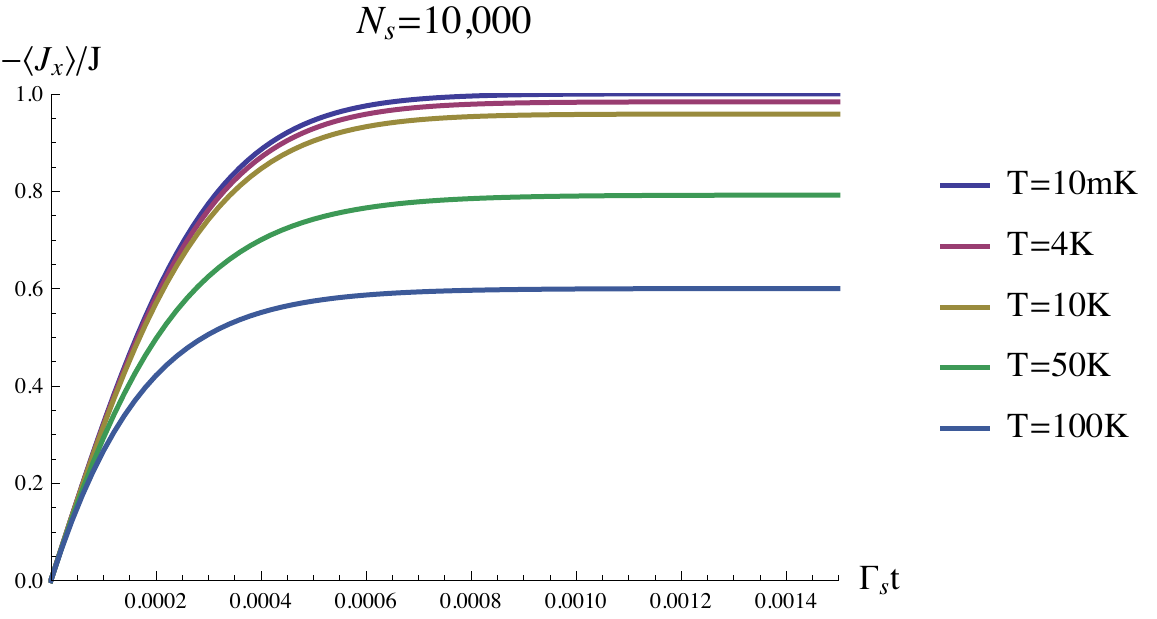}
\caption{Normalized spin expectation value $-\langle J_x\rangle/J$ of the Dicke subspace of a spin ensemble as a function of time for various equilibrium temperatures of the cavity. We consider the case of $N_s=$10, 100, 1000 and 10,000 spins in the ensemble, and a cavity with resonant frequency $\omega_c=$ 10 GHz, and a Rabi drive on resonance with the detuning $\Delta=\delta\omega-\Omega_R=0$.}
\label{fig:jxvst}
\end{center}
\end{figure}

\end{widetext}

\end{document}